# Profitable Double-Spending Attacks

Jehyuk Jang and Heung-No Lee, *Senior Member, IEEE*

*Abstract*—Our aim in this paper is to investigate the profitability of double-spending (DS) attacks that manipulate an *a priori* mined transaction in a blockchain. It was well understood that a successful DS attack is established when the proportion of computing power an attacker possesses is higher than that the honest network does. What is not yet well understood is how threatening a DS attack with less than 50% computing power used can be. Namely, DS attacks at any proportion can be of a threat as long as the chance to making a good profit exists. Profit is obtained when the revenue from making a successful DS attack is greater than the cost of carrying out one. We have developed a novel probability theory for calculating a *finite time* attack probability. This can be used to size up attack resources needed to obtain the profit. The results enable us to derive a sufficient and necessary condition on the value of a transaction targeted by a DS attack. Our result is quite surprising: we theoretically show that DS attacks at any proportion of computing power can be made profitable. Given one's transaction size, the results can also be used to assess the risk of a DS attack. An example of the attack resources is provided for the *BitcoinCash* network.

## I. INTRODUCTION

A blockchain is a distributed ledger which has originated from the desire to find a novel alternative to centralized ledgers such as transactions through third parties [1]. Besides the role as a ledger, blockchains have been applied to many areas, e.g., managing the access authority to shared data in the cloud network [2] and averting collusion in e-Auction [3]. In a blockchain network based on the proof-of-work (PoW) mechanism, each miner verifies transactions and tries to put them into a block and mold the block to an existing chain by solving a cryptographic puzzle. This series of processes is called *mining*. But the success of *mining* a block is given to only a single *miner* who solves the cryptographic puzzle for the first time. The reward of minting a certain amount of coins to the winner motivates more miners to join and remain in the network. As a result, blockchains have been designed so that the validity of transactions is confirmed by a lot of decentralized miners in the network.

A consensus mechanism is programmed for decentralized peers in a network to share a common chain. If a full-node succeeds in generating a new block, it has the latest version of the chain. All of the nodes in the network continuously communicate with each other to share the latest chain. A node may run into a situation in which it encounters mutually different chains more than one. In such a case, it utilizes a consensus rule with which it selects a single chain. Satoshi

The authors are with the School of Electrical Engineering and Computer Science, Gwangju Institute of Science and Technology (GIST), Rep. of Korea. The asterisk * indicates the corresponding author. The e-mail addresses of authors are (jjh2014@gist.ac.kr, heungno@gist.ac.kr).

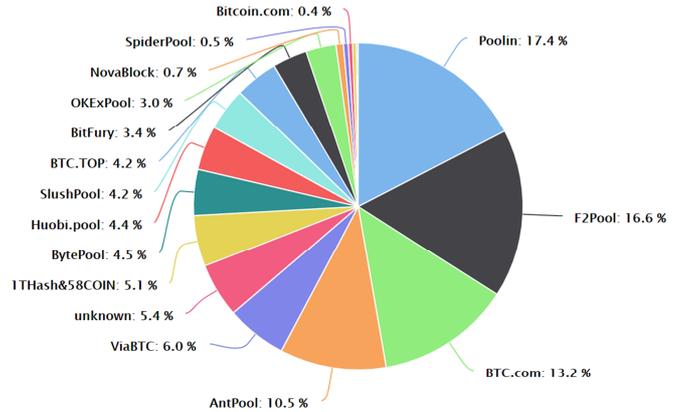

Fig. 1. Computation power distribution among the largest mining pools provided by *BTC.com* (date accessed: 5 Jan. 2020).

Nakamoto suggested the *longest chain consensus* for *Bitcoin* protocol in which the node selects the longest chain among all competing chains [1]. There are also other consensus rules [4], [5], but a common goal of consensus rules is to select the single chain by which the most computation resources have been consumed based on the belief that it may have been verified by the largest number of miners.

A double-spending (DS) attack aims to double-spend a cryptocurrency for the worth of which a corresponding delivery of goods or services has already been completed. The records of payment are written in transactions and shared in a network via the status-quo chain. Thus, to double spend, attackers need to replace the status-quo chain in the network with their new one, after taking the goods or services. For example, under the longest chain consensus, this attack will be possible if an attacker builds a longer chain than the status-quo. Nakamoto [1] and Rosenfeld [6] have shown that the higher computing power is employed, the higher probability to make a DS attack successful is. In addition, if an attacker invests more computing power than that invested by a network, a successful DS attack is guaranteed. Such attacks are called the 51% attack.

In the last few years, unfortunately, blockchain networks have been recentralized [7], [8], which make them vulnerable to DS attacks. To increase the chance of mining blocks, some nodes may form a pool of computing chips. The problem arises when a limited number of pools occupy a major proportion of the computing power in the network. For example, the pie chart shown in Fig. 1 illustrates the proportion of computing power in the Bitcoin network as of January 2020. In the chart, five pools such as Poolin, F2Pool, BTC.com, AntPool, and ViaBTC occupy more than 60% of the total computing power. In a recentralized network, since most computing resources are concentrated on a small number of pools, it could be not difficult for them to conspire to alter the block content for their own benefits, if not aiming to double spend, more probable. Indeed, there have been a number of reports in 2018 and 2019 in which cryptocurrencies such as *Verge*, *BitcoinGold*, *Ethereum Classic*, *Feathercoin*, and *Vertcoin* suffered from DS attacks and millions of US dollars have been lost [9].

In addition to the recentralization, the advent of rental services which lend the computing resources can be a concern as well [10]. Rental services such as *nicehash.com* which provide a brokerage service between the suppliers and the consumers have indeed become available. The rental service can be misused for making DS attacks easier. Indeed, *nicehash.com* attracts DS attackers to use their service by posting one-hour fees for renting 51% of the total computing power against dozens of blockchain networks on their website *crypto51.app*.

The presence of such computing resource rental services make the cost, to make a profit from double spending, significantly reduced. It is because renting a required computing power for a few hours is much cheaper than building such a computing network. The problem to consider, therefore, is to analyze the profitability of such attacks. To this end, we need to develop a new set of mathematical tools which we aim to report in this paper.

*A. Contributions*

We study the profitability of DS attacks. The concept of *cut-time* is introduced. *Cut-time* is defined to be the duration of time, from the start time to the end time of an attack. For each DS attempt, the attacker needs to pay for the cost to run his mining rig. A rational attacker would not, therefore, continue an attack indefinitely especially when operating within the regime of less than 50% computing power. To reduce the cost, the attacker needs to figure out how his attack success probability rolls out to be as the time progresses. We define that a DS attack is profitable if and only if the expected profit, the difference between revenue and cost (see equation (33)), is positive. Our contributions are summarized into two folds:

First, we theoretically show that DS attacks can be profitable not only in the regime of 51% attack but also in the sub-50% regime where the computing power invested by the attacker is smaller than that invested by the target network. Specifically, a sufficient and necessary condition is derived for profitable DS attacks on the minimum value of target transaction. In the sub-50% regime, we also show that profitable DS attacks necessitate setting a finite cut-time.

Second, we derive novel mathematical results that are useful for an analysis of the attack success time. Specifically, the probability distribution function and the first moment expectation of the attack success time have been derived. They enable us to estimate the expected profit of a DS attack for a given cut-time. All mathematical results are numerically-calculable. All the examples to find the theoretical results in this paper are provided in our web-site[1].

*B. Organization of the Paper*

In Section II, we define DS attack scenario and sufficient and necessary conditions required for successful DS attacks. Also, we define random variables that are useful in analyzing the attack profits. Section III comprises the analytic results of stochastics of the time-finite attack success. In Section IV, we define the profit function of DS attacks, followed by new theoretical results about the conditions for making them profitable. Also, an example analysis of DS attack profitability in sub-50% regime against *BitcoinCash* network is given. Section V compares our results with related works. Finally, Section VI concludes the paper with a summary.

## II. THE ATTACK MODEL

We define DS attack that we consider throughout this paper. We also define DS attack achieving (DSA) time, which is the least time spent for an occurrence of double-spending. The DSA time is a random variable derived from a random walk of Poisson counting processes (PCP).

*A. Attack Scenario*

We extend a DS attack scenario which has been considered by Nakamoto [1] and Rosenfeld [6]. Specifically, we add a time-finite attack scenario. There are two groups of miners, the normal group of honest miners and a single attacker. The normal group tends the *honest chain*.

When the attacker decides to launch a DS attack, he/she makes a target transaction for the payment of goods or services. In the target transaction, the transfer of cryptocurrency ownership from the attacker to a victim is written. We denote $t=0$ as the time at which the last block of the honest chain has been generated. At time $t=0$, the attacker announces the target transaction to normal group so that normal group starts to put it into the honest chain. At the same time $t=0$, the attacker makes a fork of the honest chain which stems from the last block and builds it in secret. We refer to this secret fork as *fraudulent chain*. In the fraudulent chain, a fraudulent transaction is contained which alters the target transaction in a way that deceives the victim and benefits the attacker.

Before shipping goods or providing services to the attacker, the victim will obviously choose to wait for a few more blocks on the honest chain in addition to the block on which the his/her transaction has been entered, i.e., so-called block confirmation. Karame *et. al.* in [11] showed the importance of block confirmation: attackers are able to double-spend against zero block-confirmation even without mining a single block on the fraudulent chain at all. The number of blocks the victim chooses to wait for is referred to as the *block confirmation number* $N_{BC} \in \mathbb{N}$, which includes the block on which the target transaction is entered.

The attacker chooses to make the fraudulent chain public if his/her attack was successful. An attack is successful if the fraudulent chain is longer than the honest chain after the moment the block confirmation is satisfied. We define two necessary conditions $\mathcal{G}^{(1)}$, $\mathcal{G}^{(2)}$ for a success of DS attack:

**Definition 1.** *A DS attack succeeds only if there exists a DS attack achieving (DSA) time $T_{DSA} \in (0, \infty)$ such that*

1. $\mathcal{G}^{(1)}$ : *(block confirmation) the length of the honest chain for the duration of time $T_{DSA}$ has grown greater than or equal to $N_{BC}$, and*

2. $\mathcal{G}^{(2)}$ : *(success in PoW competition) the length of the fraudulent chain for the duration of time $T_{DSA}$ has grown longer than that of the honest chain.*

---

[1] https://codeocean.com/capsule/2308305/tree

Rational attackers will not wait for his success indefinitely since growing the attacker's chain incurs the expense per time spent for operating the computing power. The attack thus shall put a limit to the end time to cut loss. We refer to this end time as the cut-time $t_{cut} \in \mathbb{R}^+$. A sufficient condition for the success of DS attack can be defined with applying the cut-time $t_{cut}$:

**Definition 2.** *For a given cut-time $t_{cut} \in \mathbb{R}^+$, the success of DS attack is declared if and only if there exists a DSA time $T_{DSA} \in (0, t_{cut})$ at which $\mathcal{G}^{(1)}$ and $\mathcal{G}^{(2)}$ in Definition 1 has been achieved.*

*B. Stochastic Model*

We model the conditions in Definition 2 with a stochastic model. We fit the block generation process using the PCP [12] with a given block generation rate $\lambda$ (blocks per second). Including Nakamoto [1] and Rosenfeld [6], it has been most conventional to analyze the block generation process of a blockchain using PCP. A rationale why the block generation process is modeled as PCP is given in Bowden *et. al.* [13], where experiments show the fitness of PCP model to real data samples from a live network.

We denote the lengths of the honest chain and the fraudulent chain over time $t \in (0, \infty]$ by two independent PCPs, $H(t) \in \mathbb{N}_0$ with the block generation rate $\lambda_H$ (blocks per second) and $A(t) \in \mathbb{N}_0$ with the block generation rate $\lambda_A$, respectively. Both processes start at the time origin $t = 0$ (at which the DS attack is launched) at which the both chains are at the zero states, i.e., $H(0) = A(0) = 0$. Each chain independently increases at most by 1 at a time point. An increment of 1 in the counting process occurs when the pertinent network adds a new block to its chain.

We represent the difference between $A(t)$ and $H(t)$ in a discrete-time domain as a random walk $S_i \in \mathbb{Z}$ for $i \in \mathbb{N}$. For this purpose, we first define two continuous stochastic processes $M(t)$ and $S(t)$, which are respectively defined as

$$M(t) := H(t) + A(t), \quad (1)$$

and

$$S(t) := H(t) - A(t). \quad (2)$$

The first process $M(t)$ is also a PCP [12] with the rate

$$\lambda_T := \lambda_A + \lambda_H. \quad (3)$$

The second process $S(t)$ is the continuous-time analog of the random walk $S_i \in \mathbb{Z}$ for $i \in \mathbb{N}$ such that

$$S_i := S(T_i), \quad (4)$$

where $T_i$ is the state progression time defined by

$$T_i := \inf\{t \in \mathbb{R}^+ : M(t) = i\}, \quad (5)$$

which increases as $i$ increases. Random walk $S_i$ is a stationary Markov chain starting from $S_0 = 0$. The state transition probabilities [12] are given by

$$p_A := \Pr(S_i = n - 1 \mid S_{i-1} = n) = \frac{\lambda_A}{\lambda_T}, \quad (6)$$

and

$$p_H := \Pr(S_i = n + 1 \mid S_{i-1} = n) = \frac{\lambda_H}{\lambda_T}, \quad (7)$$

for all $i \in \mathbb{N}$ and $n \in \mathbb{Z}$. The state transition probabilities $p_H$ and $p_A$ are the proportions of computing power occupied by the normal miners and that by the attacker, respectively.

We define independent and identically distributed (i.i.d.) state transition random variables $\Delta_i \in \{\pm 1\} \sim \text{Bernoulli}(p_H)$ as

$$\Delta_i := S_i - S_{i-1}, \quad (8)$$

for $i \in \mathbb{N}$. Note that $S_i = \sum_{k=0}^{i} \Delta_k$.

**Definition 3.** *A DS attack $\text{DS}(p_A, t_{cut}; N_{BC})$ is a random experiment that picks a sample $\omega \in \Omega$. Each element $\omega$ is an infinite-length sequence of pairs of $T_i$ in (5) and $\Delta_i$ in (8) for all $i \in \mathbb{N}$, i.e.,*

$$\omega := ((T_1, \Delta_1), (T_2, \Delta_2), \cdots, (T_\infty, \Delta_\infty)). \quad (9)$$

*The set $\Omega$ is the universal set of all possible $\omega$, i.e.,*

$$\Omega := \left\{ \omega \in \{\mathbb{R}^+ \times \{\pm 1\}\}^\infty \right\}. \quad (10)$$

For given a DS sample $\omega \in \Omega$ and a state index $i \in \mathbb{N}$, we denote projections

$$\pi_{T_i}(\omega) := T_i \quad (11)$$

and

$$\pi_{\Delta_i}(\omega) := \Delta_i \quad (12)$$

that retrieve the progression time $T_i$ and the transition $\Delta_i$ of the $i$-th state, respectively.

*C. DS Attack Achieving Time*

**Definition 4.** *For a DS sample $\omega$ of $\text{DS}(p_A, t_{cut}; N_{BC})$, we define the DSA time $T_{DSA}$ which measures the least one among the state progression times $\pi_{T_i}(\omega)$ of state inices $i$ at which $\omega$ achieves the necessary conditions $\mathcal{G}^{(1)}$ and $\mathcal{G}^{(2)}$ in Definition 1.*

To express $T_{DSA}$ as a random variable, we construct event sets $\mathcal{D}_j^{(1)} \subset \Omega$ and $\mathcal{D}_{i,j}^{(2)} \subset \Omega$. The sets $\mathcal{D}_j^{(1)}$ for $j \in \{N_{BC}, N_{BC}+1, \cdots, \infty\}$ consist of DS samples $\omega$ which achieves the block confirmation $\mathcal{G}^{(1)}$ at state $j$ for the first time. The sets $\mathcal{D}_{i,j}^{(2)}$ for $i \in \{j, j+1, \cdots, \infty\}$ and $j \in \{N_{BC}, N_{BC}+1, \cdots, \infty\}$ consists of $\omega$ which achieves the success in the PoW competition $\mathcal{G}^{(2)}$ at state $i$ for the first time, given that $\mathcal{G}^{(1)}$ has been already achieved at state $j$. Subsequently, we aim for the samples $\omega \in \mathcal{D}_j^{(1)} \bigcap \mathcal{D}_{i,j}^{(2)}$ to achieve the two conditions in Definition 1 at a state pair $(i, j)$ for the first time.

Formally, we first construct a set $\mathcal{D}_j^{(1)}$ focusing only on the first $j$ transitions $\Delta_k$ for $k = 1, \cdots, j$ of DS samples $\omega \in \Omega$ with two requirements; one is that they must have $N_{BC}$ number of +1's and $j - N_{BC}$ number of −1's; and the other is that the $j$-th transition $\Delta_j$ must be +1 to guarantee that they have never been achieved in any states prior to the state $j$. The former requirement implies that all $\omega \in \mathcal{D}_j^{(1)}$ hold $S_j = \sum_{k=1}^{j} \pi_{\Delta_k}(\omega) = 2N_{BC} - j$. For example, when $N_{BC} = 2$ and $j = 5$, a sequence $(+1, -1, -1, -1, +1, \cdots)$ of state transitions satisfies the first requirement, and also satisfies $S_j = 2N_{BC} - j$.

We next construct a set $\mathcal{D}_{i,j}^{(2)} \subset \Omega$ which does not care about the first $j$ transitions $\Delta_k$ for $k = 1, \cdots, j$, but only focuses on the interim transitions $\Delta_m$ for $m = j+1, \cdots, i$. By the definition, all sequences $\omega \in \mathcal{D}_{i,j}^{(2)}$ must achieve $\mathcal{G}^{(1)}$ before the $j$-th state, which implies that they must hold $S_j = 2N_{BC} - j$. The rest requirement for each $\omega \in \mathcal{D}_{i,j}^{(2)}$ is that the state changes from starting state $S_j = 2N_{BC} - j$ to state $S_i = -1$, while any interim states $S_k$ remain non-negative; i.e., $S_k \geq 0$ for each $k = j+1, \cdots, i-1$.

The sets $\mathcal{D}_j^{(1)}$ for each $j$ are mutually exclusive as each of which represents the first satisfaction of the block confirmation condition exactly at the $j$-th state. For example, if $\omega \in \mathcal{D}_5^{(1)}$ then $\omega \notin \mathcal{D}_6^{(1)}$ since $\omega$ already has achieved the block confirmation at the 5-th state for the first time before reaching the 6-th state. The sets $\mathcal{D}_{i,j}^{(2)}$ for all $(i, j)$ are also mutually exclusive for the same reason. Thus, their intersections $\mathcal{D}_j^{(1)} \bigcap \mathcal{D}_{i,j}^{(2)}$ for all $(i, j)$ are also mutually exclusive.

By Definition 4, the attack achieving time $T_{DSA}$ can be measured if there exist index pairs $(i, j)$ such that $\omega \in \mathcal{D}_j^{(1)} \bigcap \mathcal{D}_{i,j}^{(2)}$. By the mutual exclusivity of $\mathcal{D}_j^{(1)} \bigcap \mathcal{D}_{i,j}^{(2)}$, if there exists such a pair $(i, j)$, it must be unique. In addition, if $\omega \in \mathcal{D}_j^{(1)} \bigcap \mathcal{D}_{i,j}^{(2)}$, $T_{DSA}$ equals $\pi_{T_i}(\omega)$, since the state progression time $T_k$ is non-decreasing as $k$ increases. As the result, $T_{DSA}$ can be rewritten as follow,

$$T_{DSA} = \begin{cases} \pi_{T_i}(\omega), & \text{if } \exists (i, j) \in \mathbb{N}^2 : \omega \in \mathcal{D}_j^{(1)} \bigcap \mathcal{D}_{i,j}^{(2)}, \\ \infty, & \text{otherwise.} \end{cases} \quad (13)$$

## III. THE ATTACK PROBABILITIES

We aim to calculate the probability distribution function (PDF) of the DSA time $T_{DSA}$. Using this, the success probability of DS attack with a given cut-time $t_{cut}$ can be figured out as the probability that $T_{DSA} < t_{cut}$. Also, the expectation of attack success time can be calculated. The expected attack success time will be used in Section IV to estimate the attack profits.

From (13), the PDF of $T_{DSA}$ requires the probabilities of two random events; one is the state progression time $T_i$ in (5); and the other is the event that a given state index $i$ satisfies $\omega \in \mathcal{D}_j^{(1)} \bigcap \mathcal{D}_{i,j}^{(2)}$. It has been well known that $T_i$ follows Erlang distribution [12] given as

$$f_{T_i}(t) = \frac{\lambda_T (\lambda_T t)^{i-1} e^{-\lambda_T t}}{(i-1)!}. \quad (14)$$

We provide the probability for the latter event, i.e., $p_{DSA,i} = \Pr\left(\omega \in \mathcal{D}_j^{(1)} \bigcap \mathcal{D}_{i,j}^{(2)}\right)$ in the following Lemma 5.

**Lemma 5.** *For a sample $\omega$ of random experiment* $\mathrm{DS}(p_A, t_{cut}; N_{BC})$, *the probability* $p_{DSA,i} = \Pr\left(\omega \in \mathcal{D}_j^{(1)} \bigcap \mathcal{D}_{i,j}^{(2)}\right)$ *can be computed as*

$$p_{DSA,i} = \sum_{j=N_{BC}}^{j=2N_{BC}} \binom{j-1}{N_{BC}-1} C_{\frac{i-1}{2}-N_{BC}, 2N_{BC}-j} p_A^{\frac{i+1}{2}} p_H^{\frac{i-1}{2}} + \binom{i-1}{N_{BC}-1} p_H^{N_{BC}} p_A^{i-N_{BC}} \quad (15)$$

*for odd $i > 2N_{BC}$, where $C_{n,m}$ is the ballot number [14] given by*

$$C_{n,m} = \begin{cases} \frac{m+1}{n+m+1}\binom{2n+m}{n}, & n, m \in \mathbb{Z}^+ \bigcup \{0\}, \\ 0, & \text{otherwise,} \end{cases} \quad (16)$$

*and for $i \leq 2N_{BC}$ and for all even-numbered $i$, $p_{DSA,i} = 0$.*

*Proof:* See Appendix A.

By taking infinite summations of $p_{DSA,i}$ in Lemma 5 for all indices $i \in \mathbb{N}$, we can compute the probability $\mathbb{P}_{DSA}$ that a DS attack will ever achieve the necessary conditions in Definition 1.

**Corollary 6.** *For a sample $\omega$ of random experiment* $\mathrm{DS}(p_A, t_{cut}; N_{BC})$ *with $t_{cut} = \infty$, the probability $\mathbb{P}_{DSA}$ has an algebraic expression*

$$\mathbb{P}_{DSA} = \begin{cases} 1, & p_H \leq p_A, \\ 1 - p_A^{N_{BC}+1} p_H^{N_{BC}} \sum_{j=N_{BC}}^{2N_{BC}} \binom{j-1}{N_{BC}-1} A_j, & p_H > p_A, \end{cases} \quad (17)$$

where

$$A_j := p_A^{j-2N_{BC}-1} - p_H^{j-2N_{BC}-1}. \quad (18)$$

*Proof:* See Appendix B.

From (13), the PDF of $T_{DSA}$ follows the PDF of $T_i$ at a given state index $i$, if at which it holds that $\omega \in \mathcal{D}_j^{(1)} \cap \mathcal{D}_{i,j}^{(2)}$, with the probability of $p_{DSA,i}$. If there does not exist such an index $i$, with the probability of $1-\mathbb{P}_{DSA}$, then $T_{DSA} = \infty$. Thus, we can write the PDF $f_{T_{DSA}}$ of $T_{DSA}$ as follow,

$$f_{T_{DSA}}(t) = \sum_{i=2N_{BC}+1}^{\infty} p_{DSA,i} f_{T_i}(t) + (1-\mathbb{P}_{DSA})\delta(t-\infty), \quad (19)$$

where $\delta(t)$ is the Dirac delta function.

**Proposition 7.** *The PDF $f_{T_{DSA}}$ has an analytic expression:*

$$f_{T_{DSA}}(t) = \frac{p_A \lambda_T e^{-\lambda_T t} \left(p_A p_H (\lambda_T t)^2\right)^{N_{BC}}}{(2N_{BC})!}$$
$$\cdot \sum_{j=N_{BC}}^{j=2N_{BC}} \binom{j-1}{N_{BC}-1} {}_2F_3\left(\mathbf{a};\mathbf{b}; p_A p_H (\lambda_T t)^2\right) \quad (20)$$
$$+ \frac{e^{-\lambda_T t}}{t} \frac{(p_H \lambda_T t)^{N_{BC}}}{(N_{BC}-1)!} \left(e^{p_A \lambda_T t} - \sum_{i=0}^{N_{BC}} \frac{(p_A \lambda_T t)^i}{i!}\right)$$
$$+ (1-\mathbb{P}_{DSA})\delta(t-\infty),$$

where ${}_pF_q(\mathbf{a};\mathbf{b};x)$ is the generalized hypergeometric function [15] with the parameter vectors

$$\mathbf{a} = \begin{bmatrix} N_{BC}+1-j/2 \\ N_{BC}+1/2-j/2 \end{bmatrix} \quad (21)$$

and

$$\mathbf{b} = \begin{bmatrix} 2N_{BC}+2-j \\ N_{BC}+1 \\ N_{BC}+1/2 \end{bmatrix}. \quad (22)$$

*Proof:* See Appendix C.

By Definition 2, the probability $\mathbb{P}_{AS}$ that a DS attack $DS(p_A, t_{cut}; N_{BC})$ succeeds equals

$$\mathbb{P}_{AS}(t_{cut}) = \Pr(T_{DSA} < t_{cut}). \quad (23)$$

Note that for a special case of $t_{cut} = \infty$, $\mathbb{P}_{AS}(t_{cut}) = \mathbb{P}_{DSA}$, which coincides with the result in Rosenfeld [6].

It will be shown to be convenient to define the attack success time $T_{AS}$ of a DS attack as

$$T_{AS} := \begin{cases} T_{DSA}, & \text{if } T_{DSA} < t_{cut}, \\ \text{not defined}, & \text{otherwise}. \end{cases} \quad (24)$$

A random variable for $T_{DSA} > t_{cut}$ does not need to be defined since it is not useful. The PDF $f_{T_{AS}}$ of $T_{AS}$ is just a scaled version of $f_{T_{DSA}}(t)$ for $0 < t < t_{cut}$, which is given in (20), with a scaling factor of $\mathbb{P}_{AS}^{-1}$. Formally, the PDF $f_{T_{AS}}(t)$ equals

$$f_{T_{AS}}(t) = \begin{cases} \dfrac{f_{T_{DSA}}(t)}{\mathbb{P}_{AS}}, & \text{for } 0 \leq t < t_{cut}, \\ 0, & \text{for } t \geq t_{cut}. \end{cases} \quad (25)$$

The expectation of attack success time is computed as

$$\mathbb{E}_{T_{AS}}(t_{cut}) = \frac{\int_0^{t_{cut}} t f_{T_{DSA}}(t) dt}{\mathbb{P}_{AS}(t_{cut})}. \quad (26)$$

The following Proposition 8 gives an explicit formula of $\mathbb{E}_{T_{AS}}$ for the special case when $t_{cut} = \infty$.

**Proposition 8.** Let $p_M := \max(p_A, p_H)$, $p_m := \min(p_A, p_H)$. If $t_{cut} = \infty$, the expectation $\mathbb{E}_{T_{AS}}(t_{cut})$ has a closed-form expression:

$$\lim_{t_{cut} \to \infty} \mathbb{E}_{T_{AS}}(t_{cut}) = \frac{\lambda_T^{-1} \left(\sum_{j=N_{BC}}^{2N_{BC}} \binom{j-1}{N_{BC}-1} Z_j + \dfrac{N_{BC}}{p_H}\right)}{\mathbb{P}_{DSA}}, \quad (27)$$

where

$$Z_j := p_A p_m^{N_{BC}} p_M^{-(N_{BC}-j+1)} \left(\frac{2N_{BC}-2jp_m+1}{p_M-p_m}\right) \quad (28)$$
$$- j p_A^{-(N_{BC}-j)} p_H^{N_{BC}}.$$

*Proof:* See Appendix B.

IV. PROFIT OF DS ATTACK

The previous probabilistic analyses in [1] and [6] show that the success of DS attacks is not guaranteed when $p_A < 0.5$. However, DS attacks with $p_A < 0.5$ can be vigorously pursued as long as they bring profit.

*A. Profitable DS Attacks*

We analyze the *profitability* of DS attacks and to this end, we define a profit function $P$ of a DS attack $DS(C, p_A, t_{cut}; N_{BC})$, where $C$ is the value of a fraudulent transaction, in terms of revenue and operating expense (OPEX) of the computing power.

The OPEX $X$ (e.g. the rental fee for the computing power) and the block mining reward $R$ tend to increase with respect to $\lambda_A$ and the time $t$ consumed during the attack. Thus, $X$ and $R$ are expressed as functions of $\lambda_A$ and $t$, and they can be any increasing function; e.g., linear, exponential, or logarithm. We define $X$ and $R$, respectively, as follows:

$$X(\lambda_A, t) := \gamma \lambda_A t \left(\log_{x_1} x_2\right)^{\lambda_A} \left(\log_{x_3} x_4\right)^{t} \quad (29)$$

for real constants $\gamma > 0$, $x_1, x_2 > 1$, and $x_3, x_4 > 1$, and

$$R(\lambda_A, t) := \beta \lambda_A t \left(\log_{r_1} r_2\right)^{\lambda_A} \left(\log_{r_3} r_4\right)^{t} \quad (30)$$

for real constants $\beta > 0$, $r_1, r_2 > 1$, and $r_3, r_4 > 1$. We denote the ratio of $\gamma$ and $\beta$ by

$$\mu := \beta \gamma^{-1}. \quad (31)$$

With regards to $P$, if an attack succeeds, the revenue comes from $C$, as it is double-spent, and $R$ for the number of blocks minded during the time duration $T_{AS}$, i.e., $R(\lambda_A, T_{AS})$. In this case, the cost is the OPEX for the time duration $T_{AS}$, i.e., $X(\lambda_A, T_{AS})$. If the attack fails, the cost is the OPEX $X(\lambda_A, t_{cut})$ for the time duration $t_{cut}$, and there is no revenue. Hence, for a DS attack $\mathrm{DS}(C, p_A, t_{cut}; N_{BC})$, we define $P$ as follow,

$$P := \begin{cases} C + R(\lambda_A, T_{AS}) - X(\lambda_A, T_{AS}), & \text{if } T_{DSA} < t_{cut}, \\ -X(\lambda_A, t_{cut}), & \text{otherwise.} \end{cases} \quad (32)$$

Subsequently, the expected profit function is

$$\begin{aligned}\mathbb{E}_P &= \mathbb{P}_{AS}(t_{cut}) \cdot \left(C + \mathbb{E}[R(\lambda_A, T_{AS})] - \mathbb{E}[X(\lambda_A, T_{AS})]\right) \\ &\quad - (1 - \mathbb{P}_{AS}(t_{cut})) X(\lambda_A, t_{cut}) \\ &= \mathbb{P}_{AS}(t_{cut}) \cdot \left(C + \mathbb{E}[R(\lambda_A, T_{AS})]\right) - \mathbb{E}_X,\end{aligned} \quad (33)$$

where $\mathbb{E}_X$ is the expected OPEX defined as

$$\mathbb{E}_X := \mathbb{P}_{AS}(t_{cut}) \mathbb{E}[X(\lambda_A, T_{AS})] + (1 - \mathbb{P}_{AS}(t_{cut})) X(\lambda_A, t_{cut}). \quad (34)$$

**Definition 9.** *A DS attack* $\mathrm{DS}(C, p_A, t_{cut}; N_{BC})$ *is said to be profitable if and only if the expected profit* $\mathbb{E}_P > 0$, *where* $\mathbb{E}_P$ *is defined in* (33).

The key factor in determining the profitability of DS attacks is the value $C$ of the fraudulent transaction. Thus, attackers would be interested in the minimum value required for profitable DS attacks [16]. Definition 9 implies that a DS attack $\mathrm{DS}(C, p_A, t_{cut}; N_{BC})$ is profitable if and only if $C > C_{\mathrm{Req.}}$, where the required value of target transaction $C_{\mathrm{Req.}}$ is

$$C_{\mathrm{Req.}} = \frac{\mathbb{E}_X}{\mathbb{P}_{AS}} - \mathbb{E}[R(\lambda_A, T_{AS})]. \quad (35)$$

The following results in Theorem 10 and Theorem 11 focus on the case where both $X(\lambda_A, t)$ and $R(\lambda_A, t)$ are linearly increasing functions of $\lambda_A$ and $t$.

**Theorem 10.** *Suppose* $x_1 = x_2$ *and* $x_3 = x_4$ *in* (29), *and* $r_1 = r_2$ *and* $r_3 = r_4$ *in* (30). *Then, a DS attack* $\mathrm{DS}(C, p_A, t_{cut}; N_{BC})$ *for any* $p_A \in (0,1)$ *and for any* $t_{cut} \in (0, \infty]$ *is profitable if and only if* $C > C_{\mathrm{Req.}}$, *where*

$$C_{\mathrm{Req.}} = \frac{(1 - \mathbb{P}_{AS}(t_{cut}))}{\mathbb{P}_{AS}(t_{cut})} \gamma \lambda_A t_{cut} - (\mu - 1) \gamma \lambda_A \mathbb{E}_{T_{AS}}(t_{cut}). \quad (36)$$

*Proof:* Substituting $x_1 = x_2$, $x_3 = x_4$, $r_1 = r_2$, and $r_3 = r_4$ into (35) results in (36). ∎

Theorem 10 shows that not only superior attackers with $p_A \in (0.5, 1)$ but also inferior attackers with $p_A \in (0, 0.5)$ are able to expect profitable DS attacks once a high enough value $C$ greater than $C_{\mathrm{Req.}}$ of the target transaction is designed. The condition $C_{\mathrm{Req.}}$ in (36) can be pre-computed before carrying out an attack, as it stochastically estimates the future expected cost, for a given position $p_A \in (0,1)$ and a cut-time $t_{cut}$ of an attacker, and a given set of network environment parameters $\gamma$ and $\beta$.

Table I lists the resources including $C_{\mathrm{Req.}}$, $\mathbb{E}_X$, and $\mathbb{E}_{T_{AS}}$ required for profitable DS attacks using $p_A = 0.35$ and $p_A = 0.4$, when $t_{cut} = c N_{BC} \lambda_H^{-1}$ with $c = 4$. Note that the expectation of the time spent for the block confirmation equals $N_{BC} \lambda_H^{-1}$, and we let $t_{cut}$ linear to it. In other words, as normal traders wait for $N_{BC} \lambda_H^{-1}$ seconds on the average, attackers shall be tolerable as well and wait for the same scale of time duration. Note that the $\mathbb{P}_{AS}$ for $N_{BC} = 1$ is smaller than that for $N_{BC} = 3$ due to not long enough $t_{cut}$. We scaled the results by parameters $\lambda_H$ and $\gamma$, which we will explain how to obtain from internet in the next subsection.

The following Theorem 11 is for the inferior attackers with $p_A \in (0, 0.5)$ and shows the importance of setting a finite $t_{cut}$.

**Theorem 11.** *Suppose* $x_1 = x_2$ *and* $x_3 = x_4$ *in* (29), *and* $r_1 = r_2$ *and* $r_3 = r_4$ *in* (30). *Then, a DS attack* $\mathrm{DS}(C, p_A, t_{cut}; N_{BC})$ *with* $p_A \in (0, 0.5)$ *is profitable only if* $t_{cut} < \infty$.

*Proof:* For any $p_A \in (0, 0.5)$, it always holds that $\mathbb{P}_{AS} < 1$. In this case, if $t_{cut} \to \infty$ then $C_{\mathrm{Req.}} \to \infty$ from (36); i.e., infinite value $C$ of fraudulent transaction is required for a DS attack $\mathrm{DS}(C, p_A, t_{cut}; N_{BC})$ to be profitable. Thus, for a DS attack with $p_A \in (0, 0.5)$ to be profitable, a finite cut-time $t_{cut} < \infty$ must be set. ∎

Theorem 11 shows that for $p_A \in (0, 0.5)$, setting $t_{cut} = \infty$ is expected to incur infinite deficit. On the contrary, for $p_A \in (0.5, 1)$, what we have numerically checked out but omitted due to space limitation is the result that $\mathbb{E}_P$ is an increasing function of $t_{cut}$; i.e., setting $t_{cut} = \infty$ is the optimal choice in the superior attack regime. Applying $p_A \in (0.5, 1)$ and $t_{cut} = \infty$ into (36) leads to $\mathbb{P}_{AS} = 1$, and thus $C_{\text{Req.}}$ turns into

$$C_{\text{Req.}} = -(\mu - 1)\gamma \lambda_A \mathbb{E}_{T_{AS}}, \tag{37}$$

where a closed-form expression of $\mathbb{E}_{T_{AS}}$ is given in Proposition 8. In this case, if $\beta > \gamma$; i.e., $\mu > 1$, DS attacks are always profitable regardless of $C$. According to *nicehash.com*, most networks maintain $\beta > \gamma$ by the economic equilibrium. As the result, in addition to the results in [1] and [6] that DS attacks with $p_A \in (0.5, 1)$ guarantee probabilistic success, we show that such attacks guarantee economic gain as well.

### B. Profitable DS Attacks against BitcoinCash

We analyze resources required for profitable DS attacks against *BitcoinCash* network. The resources include the computing power proportion $p_A$, expected OPEX $\mathbb{E}_X$, expected attack success time $\mathbb{E}_{T_{AS}}$, and the required value of fraudulent transaction $C_{\text{Req.}}$.

To this end, we first recall the parameters involved in block mining reward $R$ and the OPEX $X$. The parameters used in (29) and (30) are assumed to $x_1 = x_2$, $x_3 = x_4$, $r_1 = r_2$, and $r_3 = r_4$ which lead to linear functions $X(\lambda_A, t)$ and $R(\lambda_A, t)$ with respect to $\lambda_A$ and $t$. There are three more parameters: $\gamma$, $\beta$, and $\lambda_H^{-1}$. From (29) and (30), the parameter $\gamma$ is the expected cost spent per generating a block; and the parameter $\beta$ is the reward per generating a block. Parameter $\lambda_H^{-1}$ is the average block generation time of the honest chain. All the parameters are different for each blockchain network.

In *BitcoinCash*, the reward $\beta$ per block mining was 12.5 BCH (without transaction fees), which is around $\beta = 0.44$ BTC per block mining (as of 26$^{\text{th}}$ Feb. 2020). The average block generation time was fixed at $\lambda_H^{-1} = 600$ seconds.

The parameter $\gamma$ is obtainable from *nicehash.com*. *BitcoinCash* uses the *SHA*-256 cryptographic puzzle for which the unit of computation is *hash*. As of 26$^{\text{th}}$ Feb. 2020, the rental fee for 1-peta (P) hashes per second for a day was around 0.017 BTC, which was around $1.97 \cdot 10^{-7}$ BTC per second. In other words, the rental fee was approximately $1.97 \cdot 10^{-22}$ BTC per the computing of a hash. Referring to *BTC.com*, the network's computing speed is 3.57-exa (E) hashes per second; i.e., $3.57\text{E} \cdot 600 = 2142\text{E}$ hashes are needed to generate one block on the average. As the result, the parameter $\gamma$ is obtained as

$$\begin{aligned}\gamma &= 1.97 \cdot 10^{-22} \, [\text{BTC/hash}] \\ &\quad \times 2142\text{E} \, [\text{hashes/block mining}] \\ &\approx 0.422 \, [\text{BTC/block mining}].\end{aligned} \tag{38}$$

Note that it holds $\beta > \gamma$. From (37), this relationship makes DS attack $DS(C, p_A, t_{cut}; N_{BC})$ with $p_A > 0.5$ and $t_{cut} = \infty$ always profitable regardless of the value $C$ of target transaction.

In case of DS attacks with $p_A < 0.5$, the cut-time $t_{cut}$ must be determined as a finite value for profitable DS attacks by Theorem 11. We set $t_{cut} = cN_{BC}\lambda_H^{-1} = 12000$ seconds with $c = 4$ and $p_A = 0.35$. We compute the resources required for profitable DS attacks against *BitcoinCash* when $N_{BC} = 5$. Results are obtainable from the values in Table I by multiplying the scaling parameters $\gamma = 0.422$ and $\lambda_H^{-1} = 600$ and by substituting $\mu = \beta\gamma^{-1} = 1.04$ and $c = 4$.

As the results, we obtain $\mathbb{P}_{AS} \approx 0.218$, $\mathbb{E}_{T_{AS}} \approx 5200$ seconds, $\mathbb{E}_X \approx 3.98$ BTC, and $C_{\text{Req.}} \approx 16.22$ BTC. One can compute expected running time; i.e., the expected time spent for a single DS attack attempt as $\mathbb{P}_{AS}\mathbb{E}_{T_{AS}} + (1 - \mathbb{P}_{AS})t_{cut}$, which is around 2 hours and 55 minutes. That is to say, attackers can repeatedly perform $n$ number of attacks every 2 hours and 55 minutes on the average. With the value $C$ of target transaction, by the strong law of large numbers, the multiple attack attempts will return net profit

TABLE I

NUMERICAL COMPUTATIONS OF REQUIRED RESOURCES FOR PROFITABLE DS ATTACKS WHEN $t_{cut} = cN_{BC}\lambda_H^{-1}$ WITH $c = 4$.

| Block Confirmation Number ($N_{BC}$) | | 1 | | 3 | | 5 | | 7 | | 9 | |
|---|---|---|---|---|---|---|---|---|---|---|---|
| Computing Power ($p_A$) | | 0.35 | 0.4 | 0.35 | 0.4 | 0.35 | 0.4 | 0.35 | 0.4 | 0.35 | 0.4 |
| AS Probability ($\mathbb{P}_{AS}$) | | 0.315 | 0.411 | 0.279 | 0.419 | 0.218 | 0.376 | 0.170 | 0.334 | 0.132 | 0.297 |
| Expected Attack Success Time ($\mathbb{E}_{T_{AS}}$) | Scaled by $\lambda_H^{-1}$ | 2.004 | 1.953 | 5.518 | 5.338 | 8.681 | 8.434 | 11.694 | 11.418 | 14.607 | 14.325 |
| Expected OPEX ($\mathbb{E}_X$) | | 1.815 | 2.106 | 5.487 | 6.139 | 9.440 | 10.436 | 13.588 | 14.977 | 17.859 | 19.716 |
| Required Value of Target Transaction ($C_{\text{Suf.}}$) | Scaled by $\gamma$ | $1.079 \cdot (1-\mu) + 4.680$ | $1.302 \cdot (1-\mu) + 3.819$ | $2.971 \cdot (1-\mu) + 16.68$ | $3.559 \cdot (1-\mu) + 11.10$ | $4.675 \cdot (1-\mu) + 38.62$ | $5.622 \cdot (1-\mu) + 22.15$ | $6.297 \cdot (1-\mu) + 73.84$ | $7.612 \cdot (1-\mu) + 37.25$ | $7.866 \cdot (1-\mu) + 127.00$ | $9.550 \cdot (1-\mu) + 56.96$ |

$n\mathbb{P}_{AS}(t_{cut}) \cdot (C - C_{Req.})$ as $n \to \infty$ with probability 1.

## V. RELATED WORKS

By Nakamoto [1] and Rosenfeld [6], the probabilities have been studied that a DS attack will ever succeed when there is no time limit, i.e., the cut-time is set to $t_{cut} = \infty$. Both of them applied PCPs to model the growth of chains $H(t)$ and $A(t)$. On one hand, the main difference between them was in probability calculations of the block confirmation process $\mathcal{G}^{(1)}$ in Definition 1. Rosenfeld applied the PCPs to both $H(t)$ and $A(t)$, whereas Nakamoto assumed the time spent for $H(t) \geq N_{BC}$ deterministic to simplify the calculation. On the other hand, they both used the gambler's ruin approach to obtain the asymptotical behavior of $S_i$ as $i \to \infty$ by manipulating the recurrence relationship between two adjacent states. Namely, their results are based on an assumption that an indefinite number of attack chances are given [17].

On the contrary, we introduce the cut-time $t_{cut}$ which generalizes analytical framework to the more interesting finite attack time and inferior attacker regime. By setting $t_{cut}$ infinite, the same result $\mathbb{P}_{DSA}$ was obtained in [6] as well. By setting a finite $t_{cut}$, our results shall be useful when attack chances are limited due to limited amount of resources such as time and cost. In addition, we show in Theorem 11 that DS attacks with $p_A < 0.5$ must set a finite $t_{cut}$ in order to expect a non-negative profit. It shall be noted that there has been no intermediate result like $p_{DSA,i}$ in Lemma 5. We use Lemma 5 to derive the novel results.

Rosenfeld [6] and Bissias et al. [18] have analyzed the profitability of DS attacks. But they put additional assumptions on the attack scenario to simplify the calculation of the attack time. Specifically, Rosenfeld assumed the attack time to be a constant. Bissias et al. assumed that the attack stops if either the normal peers or the attacker achieves the block confirmation first. On the contrary, in our model, an attack can be continued for a random attack time as long as it brings profit, even if the normal peers achieve the block confirmation before the attacker does.

In Zaghloul et. al [19], the profit of DS attack has been analyzed. Interestingly, they have discussed the need of cut-time for DS attacks with $p_A < 0.5$, which is theoretically proven in this paper in Theorem 11. They also calculated the profit of DS attacks with a finite time-limit (see Section IV-C in [19]), but their calculation was not as precise as ours in three points:

First, the probability of attack success within a finite time-limit, i.e., $\mathbb{P}_{AS}(t_{cut})$ in (23) was never considered, which requires the distribution of the DS achieving time, i.e., $T_{DSA}$ given in Proposition 7. Instead, their calculation used $\mathbb{P}_{DSA}$ referring to the result in Rosenfiled [18]. This contradicts their time-limited attack scenario, since $\mathbb{P}_{DSA}$ was resulted from the assumption of infinite time-limit.

Second, they approximated costs and revenues of DS attack spent within a time-limit. Estimation of the costs and revenues requires estimations of the numbers of blocks respectively mined by honest nodes and attackers within a time-limit, but those were assumed to be constant. This was due to the absence of the time analysis we provide in Proposition 7.

Third, they assumed the average block generation rates $\lambda_H$, $\lambda_A$ respectively by honest miners and by attackers are always the same. Since, the proportions $p_H$, $p_A$ of computing power occupied by the two groups can be quite different in general, such a result is not very useful. We agree to their assumption that most blockchains control the difficulty of block mining puzzle to keep the average speed of block generation constant, and thus $\lambda_H$ can be considered as a constant. But $\lambda_A$ should be left as a varying quantity by $p_A$. The fact is that the computing power invested by attacker cannot be monitored by the honest network and thus it cannot be reflected in the difficulty control routine.

Budish [20] conducted simulations on the profitability of DS attacks only in the cases of $p_A > 0.5$. Under the cases, a condition on the value of the target transaction that makes DS attacks not profitable has been given based on the simulations. We give theoretical and numerically-calculable results for any $p_A \in (0,1)$, which do not require massive simulations.

Gervais et.al. [21] and Sompolinsky et.al. [17] have used a Markov decision process (MDP) to analyze profits from DS attacks. These works differ from our contributions in the following regards:

First, they did not follow the DS attacks scenario considered by Nakamoto [1] and Rosenfeld [6]. Instead, the scenario in [17] was a special case of the pre-mining strategy which was introduced in [22] and [23]. We show that the success of DS attack under this scenario is even more difficult to occur than the success of the DS attack under the scenario of Nakamoto and Rosenfeld (see Appendix D for details). Also, the attack scenario in [21] went even further by modifying the condition $\mathcal{G}^{(1)}$ for block confirmation in Definition 1. Specifically, under $\mathcal{G}^{(1)}$, it is required for the honest chain to have added $N_{BC}$ blocks, while under their condition, it was the fraudulent chain to do so (see Section 3 of [21]). Thus, it was not ensured that the potential victim has shipped the goods or service, and an attack success did not guarantee for the attacker to obtain the benefit of attacking.

Second, one important new advance in this paper is the derivation of the time analysis $f_{T_{AS}}$ given in Proposition 7. When one uses the MDP framework, one can obtain similar information such as the estimations for the attack success time $\mathbb{E}_{T_{AS}}$, the future profit $P$ that an attacker will earn in the end, and the minimum value of target transaction $C_{Req.}$. But using MDP, to make such estimations, would have required extensive Monte Carlo simulations. Using our mathematical results, such estimations can be obtained without Monte Carlo simulations.

In addition, we believe that our mathematical results can be utilized in the MDP frameworks to improve the reliability of analyses. Conventionally, a rational user of an MDP will make a decision at every state whether to *stop* or to *continue* the process by comparing the rewards that will be incurred in the future by his/her decision. The rewards for *stop* actions are clear because such actions are either an attack *success* or a *give-up*. The reward for the *continue* action is complex because it needs to consider all the actions in all future possible states as well. In [17] and [21], the rewards for the *continue* action were over-simplified as they were evaluated only for the very next state and did not include the estimation of the profits in further future actions. To improve the reliability, the PDF $f_{T_{AS}}$ in Proposition 7 can be used at any intermediate Markov state to estimate the future profits. Specifically, the conditional expectation of the time left for an attack success $T_{AS}$ given $T_{AS} > \tau$ can be calculated using $f_{T_{AS}}$, where $\tau$ is the observable time elapsed for reaching the current state. Once the time left is estimated, the estimation of future profits can be updated by substituting it into (33). That is to say, at each state, the estimation of profits can be updated and used as the rewards resulting from the *continue* action.

Goffard [24] and Karame *et. al.* [25] have derived the PDFs of attack success time, but none of their DS attack scenarios matched with ours in Definition 1. In [24], Goffard derived the PDF of catch-up time spent for the fraudulent chain to catch up with the honest chain given that the length of honest chain is initially ahead by several blocks. The author used counting processes such as order statistic point process and renewal process which are more general than PCP, but there was no analytic result similar to what is given in Proposition 7. In [25], Karame *et al.* derived the PDF of the first attack success time under a fast-payment model which fixed $N_{BC} = 0$. To sum up, the attack success time in neither analysis included the time spent for achieving the first condition $\mathcal{G}^{(1)}$: the block confirmation should be realized.

## VI. CONCLUSIONS

We showed that DS attacks using 50% or a less proportion of computing power can be profitable and thus quite threatening. We provided how much quantitative resources are required to make a profitable DS attack. We derive the PDF of attack success time which enables us to figure out the operating time and the expense of mining rigs. We provided MATLAB codes on the website[2] for numerical calculation of the expected profit function in (33) and the minimum resources listed in Table I required for a profitable attack. In Table I, we summarized an example of the required resources, which is applicable to any blockchain networks by substituting $\gamma$, $\beta$, and $\lambda_H$. We also showed more specific example of the required resources against *BitcoinCash* network.

Our results quantitatively guide how to set a block confirmation number. The less the block confirmation number is, the less the minimum resource is required for a profitable attack. A solution can be utilized by the network developers to discourage such an attack. On the one hand, given a block confirmation number, we can have the value of any transaction to be set below the required value of making a profitable attack in a given network. On the other hand, given the value of transaction, the network can provide a service to inform the payee with the least block confirmation number that leads to negative DS attack profit.

---

[2] https://codeocean.com/capsule/2308305/tree

## APPENDIX A
### PROOF OF LEMMA 5

For a given sample $\omega$ and a given index $i$, the event $\omega \in \mathcal{D}_j^{(1)} \cap \mathcal{D}_{i,j}^{(2)}$ is equivalent to the event that there exists an intermediate state index $j$ such that $\omega \in \mathcal{D}_j^{(1)} \cap \mathcal{D}_{i,j}^{(2)}$. By the mutual exclusiveness of $\mathcal{D}_j^{(1)} \cap \mathcal{D}_{i,j}^{(2)}$ for integers $j$, such a state $j$ is unique if it exists. Thus, we can write the probability $p_{DSA,i}$ as follow,

$$p_{DSA,i} = \Pr\left(\exists j \in \mathbb{N}: \omega \in \mathcal{D}_j^{(1)} \cap \mathcal{D}_{i,j}^{(2)}\right)$$
$$= \sum_{j=N_{BC}}^{\infty} \Pr\left(\omega \in \mathcal{D}_j^{(1)} \cap \mathcal{D}_{i,j}^{(2)}\right). \quad (39)$$

Note that $\mathcal{D}_j^{(1)} \cap \left(\mathcal{D}_{i,j}^{(2)}\right) = \phi$ for $i \leq 2N_{BC}$, since the minimum number of states for an attack success is $2N_{BC}+1$; $N_{BC}$ number of $+1$'s state transitions for the block confirmation; and $N_{BC}+1$ number of $-1$'s state transitions for the success of PoW competition. Thus, $p_{DSA,i} = 0$ for $i \leq 2N_{BC}$.

We further explore $\mathcal{D}_j^{(1)}$ and $\mathcal{D}_{i,j}^{(2)}$. We divide the domain of state index $j$ in (39) into two exclusive domains; one is $j \leq 2N_{BC}$; and the other is $j > 2N_{BC}$. First, for $j \leq 2N_{BC}$, two sets $\mathcal{D}_j^{(1)}$ and $\mathcal{D}_{i,j}^{(2)}$ are independent, since their requirements on the state transitions are focusing on disjoint indices of state by their definitions. Formally, $\Pr\left(\omega \in \mathcal{D}_j^{(1)} \cap \mathcal{D}_{i,j}^{(2)}\right) = \Pr\left(\omega \in \mathcal{D}_j^{(1)}\right)\Pr\left(\omega \in \mathcal{D}_{i,j}^{(2)}\right)$. Second, we explore the domain $j > 2N_{BC}$. By the definition of $\mathcal{D}_j^{(1)}$, all $\omega \in \mathcal{D}_j^{(1)}$ satisfy $S_j = \sum_{k=1}^{j} \pi_{\Delta_k}(\omega) = 2N_{BC} - j$. Thus, for every $j > 2N_{BC}$, $S_j$ is already negative, which implies all $\omega \in \mathcal{D}_j^{(1)}$ satisfy both $\mathcal{G}^{(1)}$ and $\mathcal{G}^{(2)}$ at state $j$. The set $\mathcal{D}_{i,j}^{(2)} = \phi$ for $j > 2N_{BC}$ and $j < i$, since the state $S_j = 2N_{BC} - j$ contradicts one requirement of $\mathcal{D}_{i,j}^{(2)}$: the interim transitions between the states $j$ and $i$ should be non-negative. For $j > 2N_{BC}$ and $j = i$, let us set $\mathcal{D}_{i,j}^{(2)} = \Omega$, since there is no interim state to apply the requirement to. To sum up, $\mathcal{D}_j^{(1)} \cap \mathcal{D}_{i,j}^{(2)} = \mathcal{D}_i^{(1)}$ for $j > 2N_{BC}$ and $i = j$, and $\mathcal{D}_j^{(1)} \cap \left(\mathcal{D}_{i,j}^{(2)}\right) = \phi$ for $j > 2N_{BC}$ and $i > j$. Subsequently, (39) is computed as

$$p_{DSA,i} = \sum_{j=N_{BC}}^{2N_{BC}} \Pr\left(\omega \in \mathcal{D}_j^{(1)}\right)\Pr\left(\omega \in \mathcal{D}_{i,j}^{(2)}\right) + \Pr\left(\omega \in \mathcal{D}_i^{(1)}\right). \quad (40)$$

We now compute the ingredient probabilities $\Pr\left(\omega \in \mathcal{D}_j^{(1)}\right)$ and $\Pr\left(\omega \in \mathcal{D}_{i,j}^{(2)}\right)$ in (40). First, by the definition, all samples in $\mathcal{D}_j^{(1)}$ must have $N_{BC}-1$ number of $+1$'s state transitions among the first $j-1$ transitions. And the rest of the $j-1$ transitions must be valued by $-1$. In addition, the $j$-th transition must be valued by $+1$ so that the block confirmation is achieved exactly at the $j$-th state index. As the result, the probability $\Pr\left(\omega \in \mathcal{D}_j^{(1)}\right)$ equals the point mass function of a negative binomial distribution:

$$\Pr\left(\omega \in \mathcal{D}_j^{(1)}\right) = \binom{j-1}{N_{BC}-1} p_H^{N_{BC}} p_A^{j-N_{BC}}. \quad (41)$$

Second, computing the probability $\Pr\left(\omega \in \mathcal{D}_{i,j}^{(2)}\right)$ starts from counting the number of combinations of state transitions satisfying the requirements of set $\mathcal{D}_{i,j}^{(2)}$. Recall the requirements on every element of $\mathcal{D}_{i,j}^{(2)}$, for $j = N_{BC}, \cdots, 2N_{BC}$, are that the state starts from the state $S_j = 2N_{BC} - j$ and ends at the state $S_i = -1$ while all the $i-j-1$ number of interim states remain nonnegative. The $i$-th transition should be $\Delta_i = -1$ so that the success of PoW competition is achieved exactly at the state index $i$. The number of combinations of such state transitions can be counted using the ballot number $C_{n,m}$ [14], which is the number of random walks that consist of $2n+m$ steps and never become negative, starting from the origin and ending at the point $m$. In our problem, the number of random walk steps is $2n+m = i-j-1$ with $m = 2N_{BC} - j$. As a result, by multiplying the probabilities $p_A$ and $p_H$ for state transitions, the probability $\Pr\left(\omega \in \mathcal{D}_{i,j}^{(2)}\right)$ is computed as

$$\Pr\left(\omega \in \mathcal{D}_{i,j}^{(2)}\right) = C_{n,m} p_A^{(n+m+1)} p_H^{n}, \quad (42)$$

where $2n+m = i-j-1$ and $m = 2N_{BC} - j$.

Finally, substituting (41) and (42) into (40) results in (15). ∎

## APPENDIX B
### PROOF OF COROLLARY 6 AND PROPOSITION 8

#### A. Proof of Corollary 6

Taking infinite summations of $p_{DSA,i}$ for all indices $i$ results in $\mathbb{P}_{DSA}$:

$$\mathbb{P}_{DSA} = \sum_{i=2N_{BC}+1}^{\infty} p_{DSA,i} \quad (43)$$

By substituting $p_{DSA,i}$ in Lemma 5 into (43), the probability $\mathbb{P}_{DSA}$ becomes

$$\mathbb{P}_{DSA} = \sum_{j=N_{BC}}^{2N_{BC}} \binom{j-1}{N_{BC}-1} p_A \sum_{i=2N_{BC}+1}^{\infty} C_{\frac{i-1}{2}-N_{BC}, 2N_{BC}-j} \left(p_A p_H\right)^{\frac{i-1}{2}}$$
$$+ \left(\frac{p_H}{p_A}\right)^{N_{BC}} \sum_{i=2N_{BC}+1}^{\infty} \binom{i-1}{N_{BC}-1} p_A^{i}.$$

By rearranging the indices $i$ in the summations, we can obtain

$$\mathbb{P}_{DSA} = \sum_{j=N_{BC}}^{2N_{BC}} \binom{j-1}{N_{BC}-1} p_A \sum_{i=0}^{\infty} C_{i,2N_{BC}-j} (p_A p_H)^{i+N_{BC}}$$
$$+ \left(\frac{p_H}{p_A}\right)^{N_{BC}} \left( \sum_{i=N_{BC}}^{\infty} \binom{i-1}{N_{BC}-1} p_A^i - \sum_{i=N_{BC}}^{2N_{BC}} \binom{i-1}{N_{BC}-1} p_A^i \right). \quad (45)$$

We define two generating functions as

$$M_k(x) := \sum_{i=0}^{\infty} C_{i,k} x^i, \quad (46)$$

and

$$G_k(x) := \sum_{i=k}^{\infty} \binom{i}{k} x^i. \quad (47)$$

By substituting $M_k$ and $G_k$ into (45), we can write

$$\mathbb{P}_{DSA} = \sum_{j=N_{BC}}^{2N_{BC}} \binom{j-1}{N_{BC}-1} p_A (p_A p_H)^{N_{BC}} M_{2N_{BC}-j}(p_A p_H)$$
$$+ \left(\frac{p_H}{p_A}\right)^{N_{BC}} \left( p_A G_{N_{BC}-1}(p_A) - \sum_{i=N_{BC}}^{2N_{BC}} \binom{i-1}{N_{BC}-1} p_A^i \right). \quad (48)$$

The function $M_k(x)$ is a generating function of the ballot numbers $C_{i,k}$, for which the algebraic expression given in [26] is

$$M_k(x) = \left(\frac{2}{1+\sqrt{1-4x}}\right)^{k+1}. \quad (49)$$

Putting $x = p_A p_H$ into $M_k(x)$ results in

$$M_k(p_A p_H) = \left(\frac{2}{1+\sqrt{1-4p_A p_H}}\right)^{k+1}$$
$$= \begin{cases} \left(\dfrac{2}{1+\sqrt{1-4p_A(1-p_A)}}\right)^{k+1}, & p_A < p_H, \\ \left(\dfrac{2}{1+\sqrt{1-4(1-p_H)p_H}}\right)^{k+1}, & p_A \geq p_H \end{cases}$$
$$= \left(\frac{1}{p_M}\right)^{k+1}, \quad (50)$$

where $p_M := \max(p_H, p_A)$. The function $G_k(x)$ is a generating function of binomial coefficients, and the algebraic expression for it is given in [27]:

$$G_k(x) = \frac{x^k}{(1-x)^{k+1}}. \quad (51)$$

Putting $x = p_A$ into $G_k(x)$ results in

$$G_k(p_A) = p_H^{-1} \left(\frac{p_A}{p_H}\right)^k. \quad (52)$$

Substituting (50) and (52) into (48) provides

$$\mathbb{P}_{DSA} = \sum_{j=N_{BC}}^{2N_{BC}} \binom{j-1}{N_{BC}-1} p_A (p_A p_H)^{N_{BC}} p_M^{-(2N_{BC}-j+1)}$$
$$+1 - \left(\frac{p_H}{p_A}\right)^{N_{BC}} \sum_{i=N_{BC}}^{2N_{BC}} \binom{i-1}{N_{BC}-1} p_A^i. \quad (53)$$

We define $p_m := \min(p_A, p_H)$, then the relationship $p_A p_H = p_m p_M$ holds. By rearranging the order of operands, we can obtain

$$\mathbb{P}_{DSA} =$$
$$1 - \sum_{j=N_{BC}}^{2N_{BC}} \binom{j-1}{N_{BC}-1} \left( \left(\frac{p_H}{p_A}\right)^{N_{BC}} p_A^{\ j} - \frac{p_A}{p_M} \left(\frac{p_m}{p_M}\right)^{N_{BC}} p_M^{\ j} \right), \quad (54)$$

which is equal to (17). ∎

### B. Proof of Proposition 8

From (19) and (26), when $t_{cut} = \infty$, we obtain

$$\mathbb{E}_{T_{AS}} = \frac{\lim_{t_{cut} \to \infty} \int_0^{t_{cut}} t f_{T_{DSA}}(t) dt}{\mathbb{P}_{AS}(t_{cut})}$$
$$= \frac{\sum_{i=2N_{BC}+1}^{\infty} \mathbb{E}[T_i] p_{DSA,i}}{\mathbb{P}_{DSA}} \quad (55)$$
$$= \frac{\sum_{i=2N_{BC}+1}^{\infty} \frac{i}{\lambda_T} p_{DSA,i}}{\mathbb{P}_{DSA}},$$

where $E[T_i] = i\lambda_T^{-1}$ [12]. By substituting $p_{DSA,i}$ in (15) into (55) and rearranging the order of operands, we obtain

$$\lambda_T \mathbb{P}_{DSA} \mathbb{E}_{T_{AS}} = \sum_{j=N_{BC}}^{2N_{BC}} \binom{j-1}{N_{BC}-1} \sum_{i=2N_{BC}}^{\infty} (i+1) C_{\frac{i}{2}-N_{BC},2N_{BC}-j} p_A^{\frac{i+2}{2}} p_H^{\frac{i}{2}}$$
$$+ \sum_{i=N_{BC}-1}^{\infty} (i+1) \binom{i}{N_{BC}-1} p_A^{i+1-N_{BC}} p_H^{N_{BC}}$$
$$- \sum_{i=N_{BC}-1}^{2N_{BC}-1} (i+1) \binom{i}{N_{BC}-1} p_A^{i+1-N_{BC}} p_H^{N_{BC}}. \quad (56)$$

By rearranging the indices of summations, we arrive at

$$\lambda_T \mathbb{P}_{DSA} \mathbb{E}_{T_{AS}} = \sum_{j=N_{BC}}^{2N_{BC}} \binom{j-1}{N_{BC}-1} p_A^{N_{BC}+1} p_H^{N_{BC}}$$
$$\cdot \sum_{i=0}^{\infty} (2i+2N_{BC}+1) C_{i,2N_{BC}-j} (p_A p_H)^i$$
$$+ p_A \left(\frac{p_H}{p_A}\right)^{N_{BC}} \sum_{i=N_{BC}-1}^{\infty} (i+1) \binom{i}{N_{BC}-1} p_A^i \quad (57)$$
$$- \sum_{i=N_{BC}}^{2N_{BC}} i \binom{i-1}{N_{BC}-1} p_A^{i-N_{BC}} p_H^{N_{BC}}.$$

By substituting the generating functions $M_k(x)$ and $G_k(x)$ defined respectively in (46) and (47), (57) becomes

$$\lambda_T \mathbb{P}_{DSA} \mathbb{E}_{T_{AS}} = \sum_{j=N_{BC}}^{2N_{BC}} \binom{j-1}{N_{BC}-1} p_A^{N_{BC}+1} p_H^{N_{BC}}$$
$$\cdot \left( 2 \sum_{i=0}^{\infty} i C_{i,2N_{BC}-j} (p_A p_H)^i + (2N_{BC}+1) M_{2N_{BC}-j}(p_A p_H) \right)$$
$$+ p_A \left(\frac{p_H}{p_A}\right)^{N_{BC}} \left( \sum_{i=N_{BC}-1}^{\infty} i \binom{i}{N_{BC}-1} p_A^i + G_{N_{BC}-1}(p_A) \right)$$
$$- \sum_{i=N_{BC}}^{2N_{BC}} i \binom{i-1}{N_{BC}-1} p_A^{i-N_{BC}} p_H^{N_{BC}}. \quad (58)$$

We use the following relationships,

$$\sum_{i=0}^{\infty} i C_{i,k} x^i = x M_k'(x) \tag{59}$$

and

$$\sum_{i=k}^{\infty} i \binom{i}{k} x^i = x G_k'(x), \tag{60}$$

and their derivatives are given by

$$\begin{aligned} M_k'(x) &:= \frac{d}{dx} M_k(x) \\ &= \sum_{i=0}^{\infty} i C_{i,k} x^{i-1} \\ &= \frac{(k+1)}{\sqrt{1-4x}} \left( \frac{2}{1+\sqrt{1-4x}} \right)^{k+2} \end{aligned} \tag{61}$$

and

$$\begin{aligned} G_k'(x) &:= \frac{d}{dx} G_k(x) \\ &= \sum_{i=k}^{\infty} i \binom{i}{k} x^{i-1} \\ &= \frac{(k x^{k-1} + x^k)}{(1-x)^{k+2}}. \end{aligned} \tag{62}$$

By substituting (59) and (60) into (58), we obtain

$$\begin{aligned} \lambda_T \mathbb{P}_{DSA} \mathbb{E}_{T_{AS}} &= \\ &\sum_{j=N_{BC}}^{2N_{BC}} \binom{j-1}{N_{BC}-1} p_A^{N_{BC}+1} p_H^{N_{BC}} \\ &\cdot \left( 2 p_A p_H M_{2N_{BC}-j}'(p_A p_H) + (2N_{BC}+1) M_{2N_{BC}-j}(p_A p_H) \right) \\ &+ p_A \left( \frac{p_H}{p_A} \right)^{N_{BC}} \left( p_A G_{N_{BC}-1}'(p_A) + G_{N_{BC}-1}(p_A) \right) \\ &- \sum_{i=N_{BC}}^{2N_{BC}} i \binom{i-1}{N_{BC}-1} p_A^{i-N_{BC}} p_H^{N_{BC}} \end{aligned} \tag{63}$$

Putting $x = p_A p_H$ into $M_k'(x)$ in (61) results in

$$\begin{aligned} M_k'(p_A p_H) &= M_k'(p_m p_M) \\ &= \frac{(k+1)}{1 - 2 p_m} \left( \frac{1}{p_M} \right)^{k+2}. \end{aligned} \tag{64}$$

Putting $x = p_A$ into $G_k'(x)$ in (62) gives

$$G_k'(p_A) = \frac{(k p_A^{k-1} + p_A^k)}{p_H^{k+2}}. \tag{65}$$

By substituting (50), (52), (64), and (65) into (63), we finally obtain (27). ∎

## APPENDIX C
### PROOF OF PROPOSITION 7

We use a generating function and generalized hypergeometric functions to compute the infinite summations in (19).

By substituting $p_{DSA,i}$ in (15) and $f_{T_i}(t)$ in (14) into (19), we arrive at

$$\begin{aligned} f_{T_{DSA}}(t) &- (1 - \mathbb{P}_{DSA}) \delta(t - \infty) = \\ &\sum_{j=N_{BC}}^{j=2N_{BC}} \binom{j-1}{N_{BC}-1} \sum_{i=2N_{BC}+1}^{\infty} C_{\frac{i-1}{2} - N_{BC}, 2N_{BC} - j} p_A^{\frac{i+1}{2}} p_H^{\frac{i-1}{2}} \frac{\lambda_T^i t^{i-1} e^{-\lambda_T t}}{(i-1)!} \\ &+ \sum_{i=2N_{BC}+1}^{\infty} \binom{i-1}{N_{BC}-1} p_H^{N_{BC}} p_A^{i-N_{BC}} \frac{\lambda_T^i t^{i-1} e^{-\lambda_T t}}{(i-1)!}. \end{aligned} \tag{66}$$

By rearranging the indices of summations and the order of operands, we obtain

$$\begin{aligned} f_{T_{DSA}}(t) &- (1 - \mathbb{P}_{DSA}) \delta(t - \infty) = \\ &\sum_{j=N_{BC}}^{j=2N_{BC}} \binom{j-1}{N_{BC}-1} \sum_{i=0}^{\infty} \Bigg( C_{i, 2N_{BC}-j} p_A^{N_{BC}+i+1} p_H^{N_{BC}+i} \\ &\qquad \cdot \frac{\lambda_T^{2N_{BC}+2i+1} t^{2N_{BC}+2i} e^{-\lambda_T t}}{(2N_{BC}+2i)!} \Bigg) \\ &+ \left( \frac{p_H}{p_A} \right)^{N_{BC}} e^{-\lambda_T t} \Bigg( \sum_{i=N_{BC}}^{\infty} \binom{i-1}{N_{BC}-1} p_A^i \frac{\lambda_T^i t^{i-1}}{(i-1)!} \\ &\qquad - \sum_{i=N_{BC}}^{2N_{BC}} \binom{i-1}{N_{BC}-1} p_A^i \frac{\lambda_T^i t^{i-1}}{(i-1)!} \Bigg). \end{aligned} \tag{67}$$

We can define two generating functions as

$$\begin{aligned} B(x) &:= \sum_{i=0}^{\infty} C_{i, 2N_{BC}-j} \frac{x^i}{(2N_{BC}+2i)!} \\ &= (2N_{BC} - j + 1) \sum_{i=0}^{\infty} \frac{(2i + 2N_{BC} - j)!}{i!(i + 2N_{BC} - j + 1)!} \frac{x^i}{(2N_{BC}+2i)!}, \end{aligned} \tag{68}$$

and

$$\begin{aligned} H(x) &:= \sum_{i=N_{BC}}^{\infty} \binom{i-1}{N_{BC}-1} \frac{x^{i-1}}{(i-1)!} \\ &= \sum_{i=N_{BC}-1}^{\infty} \binom{i}{N_{BC}-1} \frac{x^i}{i!}. \end{aligned} \tag{69}$$

By substituting $B(x)$ and $H(x)$ into (67), we obtain

$$\begin{aligned} f_{T_{DSA}}(t) &- (1 - \mathbb{P}_{DSA}) \delta(t - \infty) = \\ &\sum_{j=N_{BC}}^{j=2N_{BC}} \binom{j-1}{N_{BC}-1} p_A \lambda_T e^{-\lambda_T t} \left( p_A p_H (\lambda_T t)^2 \right)^{N_{BC}} B\left( p_A p_H (\lambda_T t)^2 \right) \\ &+ \left( \frac{p_H}{p_A} \right)^{N_{BC}} e^{-\lambda_T t} \Bigg( p_A \lambda_T H(p_A \lambda_T t) - \sum_{i=N_{BC}}^{2N_{BC}} \binom{i-1}{N_{BC}-1} p_A^i \frac{\lambda_T^i t^{i-1}}{(i-1)!} \Bigg). \end{aligned} \tag{70}$$

We replace function $B(x)$ in (68) with the generalized hypergeometric functions. For this purpose, we first denote the sequences in $B(x)$ by

$$\beta_i := \frac{(2i + 2N_{BC} - j)!}{i!(i + 2N_{BC} - j + 1)!} \frac{1}{(2N_{BC}+2i)!}, \tag{71}$$

and

$$\beta_0 := \frac{1}{(2N_{BC} - j + 1)(2N_{BC})!}. \tag{72}$$

Next, the function $B(x)$ can be rewritten as

$$B(x) = (2N_{BC} - j + 1)\sum_{i=0}^{\infty} \beta_i x^i$$
$$= (2N_{BC} - j + 1)\beta_0 \left(x^0 + \frac{\beta_1}{\beta_0}x^1 + \frac{\beta_2}{\beta_1}\frac{\beta_1}{\beta_0}x^2 + \cdots\right). \quad (73)$$

The reformulated sequence in (73) is computed by

$$\frac{\beta_{i+1}}{\beta_i} = \frac{(i+1+N_{BC}-j/2)(i+1/2+N_{BC}-j/2)}{(i+2+2N_{BC}-j)(i+1+N_{BC})(i+1/2+N_{BC})(i+1)}, \quad (74)$$

which has 2 polynomials in $i$ on the numerator and 3 polynomials in $i$ except for $(i+1)$ on the denominator. $B(x)$ can be expressed in terms of a generalized hypergeometric function ${}_2F_3$ [15] as follow,

$$B(x) = (2N_{BC} - j + 1)\beta_0 \, {}_2F_3(\mathbf{a}_j; \mathbf{b}_j; x)$$
$$= \frac{1}{(2N_{BC})!} {}_2F_3(\mathbf{a}_j; \mathbf{b}_j; x), \quad (75)$$

where vectors $\mathbf{a}_j$ and $\mathbf{b}_j$ respectively defined in (21) and (22) are the constants in the polynomials in $i$ of the numerator and denominator in (74), respectively.

We use a closed-form expression of generating function $H(x)$ in (69) given by

$$H(x) = \sum_{i=N_{BC}-1}^{\infty}\binom{i}{N_{BC}-1}\frac{x^i}{i!}$$
$$= \frac{1}{(N_{BC}-1)!}\sum_{i=N_{BC}-1}^{\infty}\frac{x^i}{(i-N_{BC}+1)!} \quad (76)$$
$$= \frac{x^{N_{BC}-1}}{(N_{BC}-1)!}e^x,$$

where the following relationship is used [28]:

$$\sum_{i=0}^{\infty}\frac{x^i}{i!} = e^x. \quad (77)$$

By substituting (75) and (76) into (70), we arrive at

$$f_{T_{DSA}}(t) - (1-\mathbb{P}_{DSA})\delta(t-\infty) = \frac{p_A\lambda_T e^{-\lambda_T t}\left(p_A p_H(\lambda_T t)^2\right)^{N_{BC}}}{(2N_{BC})!}$$
$$\cdot \sum_{j=N_{BC}}^{j=2N_{BC}}\binom{j-1}{N_{BC}-1}{}_2F_3(\mathbf{a}_j; \mathbf{b}_j; p_A p_H(\lambda_T t)^2)$$
$$+\left(\frac{p_H}{p_A}\right)^{N_{BC}}e^{-\lambda_T t}\left(p_A\lambda_T\frac{(p_A\lambda_T t)^{N_{BC}-1}}{(N_{BC}-1)!}e^{p_A\lambda_T t}\right.$$
$$\left. -\sum_{i=N_{BC}}^{2N_{BC}}\binom{i-1}{N_{BC}-1}p_A^i\frac{\lambda_T^i t^{i-1}}{(i-1)!}\right)$$
$$= \frac{p_A\lambda_T e^{-\lambda_T t}\left(p_A p_H(\lambda_T t)^2\right)^{N_{BC}}}{(2N_{BC})!}$$
$$\cdot \sum_{j=N_{BC}}^{j=2N_{BC}}\binom{j-1}{N_{BC}-1}{}_2F_3(\mathbf{a}_j; \mathbf{b}_j; p_A p_H(\lambda_T t)^2)$$
$$+\left(\frac{p_H}{p_A}\right)^{N_{BC}}e^{-\lambda_T t}\left(p_A\lambda_T\frac{(p_A\lambda_T t)^{N_{BC}-1}}{(N_{BC}-1)!}e^{p_A\lambda_T t}\right.$$
$$\left. -\frac{1}{(N_{BC}-1)!}\sum_{i=N_{BC}}^{2N_{BC}}p_A^i\frac{\lambda_T^i t^{i-1}}{(i-N_{BC})!}\right). \quad (78)$$

We obtain (20) by rearranging the indices of the summations and the order of operands. ∎

## APPENDIX D
## COMPARISON OF ATTACK SUCCESS PROBABILITY WITH [17]

In [17], a different DS success condition other than the conditions in Definition 1 has been used. Specifically, the only condition was to have the fraudulent chain to grow longer than the honest chain by $N_{BC}$, i.e., $A(t) > H(t) + N_{BC}$ (see Section 7 of [17]). We refer to $\mathbb{P}_{\text{pre-mine}}$ as the probability of satisfying this condition. The literature has shown that satisfying this condition suffices a success of DS attack [17]. What they have not shown, however, is that this condition is not a necessary one. Thus, we here aim to show that their condition is indeed not a necessary condition, by showing that $\mathbb{P}_{DSA} > \mathbb{P}_{\text{pre-mine}}$ for all $p_A \in (0, 0.5)$. First, it has been given that $\mathbb{P}_{\text{pre-mine}} = (p_A/p_H)^{N_{BC}+1}$. Under the condition of [17], it is required that the fraudulent chain catches up with the honest chain with additional $N_{BC}$ blocks. The catch-up probability has been derived by Nakamoto in [1] using the gambler's ruin approach as $(p_A/p_H)^k$, where $k$ is the number of blocks that the honest chain leads by at the initial status. Next, we refer to an intermediate step in the derivation of $\mathbb{P}_{DSA}$ by Rosenfeld [6]:

$$\mathbb{P}_{DSA} = \sum_{k=0}^{N_{BC}+1}\binom{N_{BC}+k-1}{k}p_H^{N_{BC}}p_A^k\left(\frac{p_A}{p_H}\right)^{N_{BC}-k+1}$$
$$+ \sum_{k=N_{BC}+2}^{\infty}\binom{N_{BC}+k-1}{k}p_H^{N_{BC}}p_A^k. \quad (79)$$

Finally, clear inequalities can be used to show $\mathbb{P}_{DSA} > \mathbb{P}_{\text{pre-mine}}$:

$$\mathbb{P}_{DSA} > \sum_{k=0}^{N_{BC}+1}\binom{N_{BC}+k-1}{k}p_H^{N_{BC}}p_A^k\left(\frac{p_A}{p_H}\right)^{N_{BC}-k+1}$$
$$+ \sum_{k=N_{BC}+2}^{\infty}\binom{N_{BC}+k-1}{k}p_H^{N_{BC}}p_A^k\left(\frac{p_A}{p_H}\right)^{N_{BC}+1}$$
$$> \left(\frac{p_A}{p_H}\right)^{N_{BC}+1}\sum_{k=0}^{\infty}\binom{N_{BC}+k-1}{k}p_H^{N_{BC}}p_A^k$$
$$= \left(\frac{p_A}{p_H}\right)^{N_{BC}+1} = \mathbb{P}_{\text{pre-mine}}. \quad (80)$$

For numerical example, when $p_A = 0.35$ and $N_{BC} = 5$ the probabilities can be computed as $\mathbb{P}_{DSA} = 0.2287$ and $\mathbb{P}_{\text{pre-mine}} = 0.0244$. As the gap is significant, it is shown that the DS attack success condition defined in [17] was indeed only a sufficient condition, set to be too strict.